# Additional Exoplanet Science Enabled by FINESSE

Robert T. Zellem[1], Jonathan J. Fortney[2], Mark R. Swain[1], Geoffrey Bryden[1], John W. Chapman[1], Nicolas B. Cowan[3], Tiffany Kataria[1], Laura Kreidberg[4,5], Michael R. Line[6], Julianne I. Moses[7], Vivien Parmentier[8], Gael M. Roudier[1], and Kevin B. Stevenson[9]

[1]Jet Propulsion Laboratory, California Institute of Technology, 4800 Oak Grove Drive, Pasadena, CA 91109, USA; rzellem@jpl.nasa.gov; 1-626-379-9418
[2]Department of Astronomy & Astrophysics, University of California, Santa Cruz, CA, USA
[3]McGill Space Institute, 3550 University St, #202, Montréal, QC H3A 2A7, Canada
[4]Harvard Society of Fellows, 78 Mt. Auburn St., Cambridge, MA 02138, USA
[5]Harvard-Smithsonian Center for Astrophysics, 60 Garden St., Cambridge, MA 02138
[6]School of Earth & Space Exploration, Arizona State University, Tempe AZ 85287, USA
[7]Space Science Institute, 4750 Walnut Street, Suite 205, Boulder, CO, 80301, USA
[8]Aix Marseille Univ, CNRS, LAM, Laboratoire d'Astrophysique de Marseille, Marseille, France
[9]Space Telescope Science Institute, Baltimore, MD 21218, USA

# 1. An Introduction to FINESSE

The Fast Infrared Exoplanet Spectroscopy Survey Explorer (FINESSE) mission, recently selected by NASA's Explorer program to proceed to Step 2 study, would conduct a large-scale, uniform survey of exoplanet atmospheres and create a statistically significant sample for comparative planetology studies. The technical capabilities of the FINESSE mission are driven by a focused scientific program to study planetary formation and climate mechanisms. In addition, its mission capability enables a broad range of exoplanet science topics including exploring the role of non-equilibrium chemistry, atmospheric evolution, and the star-planet connection (Fig. 1). FINESSE's survey capability would enable follow-up exploration of TESS discoveries and provide a broader context for interpreting detailed JWST observations.

The FINESSE mission design is driven by the objectives of measuring key physical parameters that determine how planets form and what establishes planetary climate. To achieve these objectives, FINESSE would measure the mass-metallicity relationship, the C/O ratio, albedo, heat redistribution, presence of aerosols (clouds and hazes), and other atmospheric properties [1-11]. To meet these requirements, FINESSE would measure transit spectra of at least 500 planets and phase-resolved emission spectra of 100 planets (Figs. 2 and 3) using a 75 cm telescope coupled to a 0.5-5.0 µm spectrometer. FINESSE's wavelength range, which expands the upon Hubble/WFC3's spectral coverage by a factor of 7.5 (Fig. 3), is optimized for detection and accurate spectral retrieval of the key molecular and aerosol opacity sources in exoplanet atmospheres, as identified by a previous study [14]. FINESSE features high pointing stability, using a single detector for fine guidance and science, and a stable thermal design with observations from L2 halo orbit, allowing it to access up to half of the sky at any given moment in time, with a full sky field of regard every four months.

The FINESSE baseline target sample includes 500 known and TESS-predicted [15] transiting exoplanets chosen according to a figure of merit that ranks them by the predicted precision of measuring 1 scale height of atmospheric spectral modulation [16; Eqn. 10]. Emission targets are similarly selected via a ranking system according to the estimated precision of measuring their eclipse depth at 2.25 µm (the middle of the FINESSE bandpass):

$$FOM_{eclipse} = \frac{F_p R_p^2 F_s^{-1} R_s^{-2}}{10^{0.2 H-mag}}$$

where $F_p$ and $F_s$ are the fluxes of the planet and star, respectively, $R_p$ and $R_s$ are the radii of the planet and star, respectively, and H-mag is the host star's apparent magnitude in the H-band. Note that both of these figures of merits are agnostic of the observing platform and thus can be used for any telescope.

Developing the explicit flow down of FINESSE's requirements is part of a Concept Study Report that the FINESSE team is preparing for the NASA Explorer Program. Given the technical capabilities of the FINESSE mission concept that are needed to meet the science requirements, the purpose of this white paper is to explore and highlight the broader discovery potential of FINESSE.

# 2. Background

Transit spectroscopy has achieved a number of breakthroughs in the characterization of exoplanet atmospheres over the last 15 years [17-20], with Hubble, Spitzer, and ground-based telescopes detecting ~50 exoplanet atmospheres. However, except in a handful of exceptionally favorable cases, the data obtained so far cannot be used to determine the fundamental characteristics of the planets' atmospheres due to the lack of the spectral resolution and wavelength coverage necessary to break the degeneracies in theoretical models. This problem would be further exacerbated with the launch of TESS, which would discover thousands of new worlds [15]. While JWST would provide detailed characterization of individual exoplanet atmospheres, it is not suited for conducting a comprehensive, uniform, and statistically significant survey of bright transiting systems because it has a limited per-observation spectral bandpass and insufficient time to observe hundreds of planets [21]. Indeed, a community effort concluded that "*a dedicated mission is needed to … perform a uniform atmospheric survey for hundreds of the brightest TESS planets*" [21].

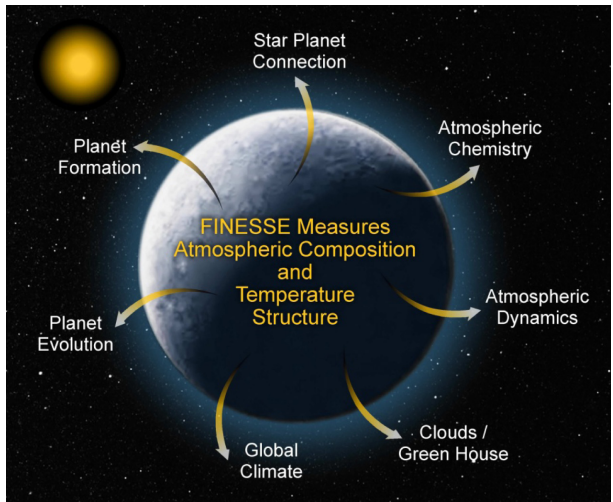

**Figure 1.** FINESSE, via its spectroscopic measurements of transits and phase-resolved emission, would provide key measurements that would advance our understanding of planetary formation and evolution, global climate, and atmospheric chemistry and physics under conditions of strong stellar forcing.

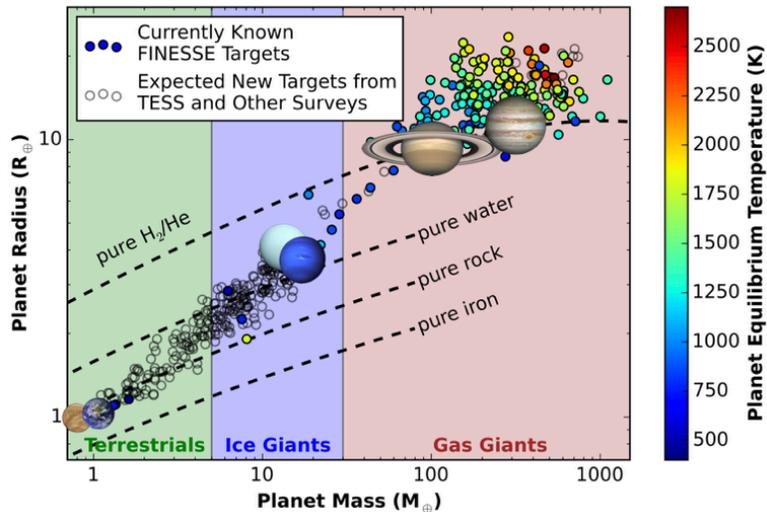

**Figure 2. FINESSE would determine atmospheric compositions, thermal structures, energy budgets, and dynamics of a statistically significant sample of diverse worlds.**

### 3. Additional Science Enabled by FINESSE: Planetary Discovery Space

The planets targeted by FINESSE are much more diverse than the planets in our Solar System. We know little about these mysterious worlds that orbit so close to their parent stars. Some of the planets that FINESSE would target could have solid surfaces with outgassed "secondary" atmospheres, some could be "water worlds" containing a large bulk fraction of water by mass, some could have exposed magma on their surfaces or even vast magma oceans, and some could be so large and hot that their atmospheric properties resemble stars more than our Solar System giant planets. Unanticipated discoveries are likely and could revolve around the general themes of "meet the planetary family" and "how atmospheres work."

The FINESSE survey would allow us to place our own Solar System into a broader context of the diversity of planets in the Galaxy. Under the general theme of "meet the planetary family," even the most unusual or outlier planets targeted by FINESSE would have an important story to tell us. In terms of the general theme of "how atmospheres work," FINESSE opens the exciting possibility of new discoveries in atmospheric chemistry and evolution, as is described below.

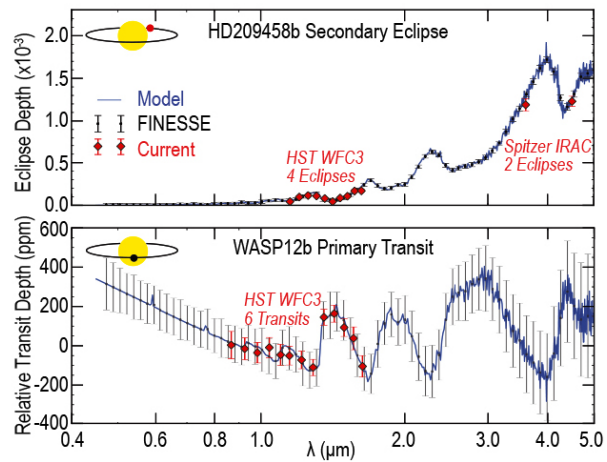

**Figure 3. FINESSE would reveal the essential properties of exoplanet atmospheres using emission (top) and transmission (bottom) spectroscopy over a wide spectral range.** HST and Spitzer measurements [12, 13] for typical FINESSE targets (red) compared to model spectra with FINESSE ±1-σ measurement uncertainty envelopes for a single transit/eclipse (gray). FINESSE has wider, instantaneous spectral coverage compared to HST and Spitzer.

**Table 1.** Molecules and locations of their prominent bands to be targeted by FINESSE.

|  | Molecule | Molecular Bands (μm) |
|---|---|---|
| Key Molecules | $H_2O$ | 0.82, 0.94, 1.13, 1.38, 1.9, 2.69 |
|  | $CH_4$ | 0.79, 0.86, 1.65, 2.2, 2.31, 2.37, 3.3 |
|  | $CO_2$ | 1.21, 1.57, 1.6, 2.03, 4.25 |
|  | CO | 1.57, 2.35, 4.7 |
| Additional Possible Chemical Species | $C_2H_2$ | 1.52, 3.0 |
|  | HCN | 3.0 |
|  | $NH_3$ | 0.93, 1.5, 2, 2.25, 2.9, 3.0 |
|  | $C_2H_4$ | 3.22, 3.34 |
|  | $C_2H_6$ | 1.95, 2.3, 2.6, 2.95, 3.07, 3.34 |
|  | $H_2S$ | 2.5, 3.8 |
|  | Na | 0.59 |
|  | K | 0.77 |
|  | TiO | 0.7–3.0, 3.0–3.5 |
|  | VO | 0.7–2.5 |

### 3.1 Disequilibrium Chemistry

The FINESSE exoplanet survey is a laboratory to study the influence of extreme versions of the disequilibrium processes that exist in temperate atmospheres. FINESSE would greatly expand our understanding of exoplanet atmospheric chemistry through the detection of numerous gaseous constituents, many of which probe disequilibrium chemistry (Table 1). Atmospheric composition is often assumed to be in thermochemical equilibrium in existing exoplanet radiative/dynamical models. In such a state, the constituent abundances depend only on temperature, pressure, and bulk elemental abundances. The atmospheric composition then varies dramatically with altitude, latitude, and longitude as a result of the significant changes in the 3D atmospheric thermal structure on the close-in, transiting, FINESSE target planets [10, 22]. For example, at infrared photospheric pressures, CO is expected to be the dominant carrier of carbon on the dayside of many hot Jupiters, whereas $CH_4$ would dominate on the terminators and night side. For very hot Jupiters, metal-bearing species such as TiO, VO, Na, K, or FeH could be present in the gas phase on the dayside, while the metals and silicates would be tied up in condensates on the cooler night side. The resulting predicted dramatic changes in three-dimensional composition would be detectable by FINESSE and have important implications for the atmosphere as a whole, given that the composition feeds back to control the radiative properties, and thus temperatures and winds within the atmosphere.

However, rapid atmospheric transport in the presence of slower chemical kinetics can cause the composition to "quench" out of equilibrium, homogenizing the atmospheric composition both vertically and horizontally [11, 22-24]. Models that consider disequilibrium quenching have very different predictions with regard to the spatial distribution of atmospheric constituents. Moreover, the strong UV irradiation received by the planet can also break apart molecules, leading to the formation of new photochemical constituents, such as HCN, $C_2H_2$, other hydrocarbons, and high-altitude hazes [11, 23, 25-30]. Such photochemistry alters the expected equilibrium composition in the atmospheres of all but the hottest exoplanets (Fig. 4). Retrieved abundances from either transmission or longitudinally-resolved emission spectra would reveal the degree to which disequilibrium chemical processes, such as transport-induced quenching and photochemistry, perturb the composition away from chemical equilibrium [11, 23]. The constraints provided by FINESSE would elucidate the relative roles of equilibrium and disequilibrium chemistry and be of critical importance for furthering our understanding of the coupling between chemistry, climate, aerosol formation, and dynamics in exoplanet atmospheres.

More generally, FINESSE would provide the large sample size needed for a statistically significant survey into how the bulk atmospheric composition and aerosol properties vary with other characteristics of a planet and its host star. Our understanding of how planets form and what processes control their diversity is at its infancy. FINESSE observations would advance the field by providing the data needed to identify and understand how bulk atmospheric properties, such as C/O ratio and metallicity, correlate with a planet's mass, radius, orbital properties, incident radiation levels, and its host star's mass, spectral type, effective temperature, age, activity level, and composition. The identification of such correlations would provide clues to formation and evolutionary processes that spawned this diversity and would launch the field of comparative exoplanetology. Large, uniform statistical studies for comparative planetology cannot be performed effectively from a multi-use platform such as JWST and require a dedicated mission such as FINESSE [21].

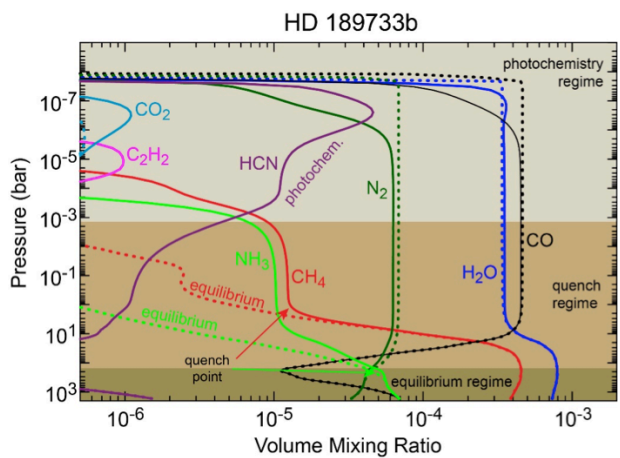

**Figure 4.** Disequilibrium processes, e.g., quenching and photochemistry, are predicted to strongly alter the molecular mixing ratios in exoplanet atmospheres [11]. **FINESSE, with its broad, instantaneous spectral coverage and diverse, statistical sample would probe disequilibrium processes for these molecules across a wide range of planetary types.**

Ultimately, detection of life outside of our Solar System is expected to rely on identification of non-equilibrium chemical species. By detecting the frequency and likely origin of non-equilibrium conditions, FINESSE data would provide an important context for future missions targeting potentially habitable worlds.

### 3.2 Atmospheric Evolution

The star's high-energy radiation affects a planet's atmospheric evolution and disequilibrium species. The host star's UV flux modifies the planet's atmospheric chemistry, while the extreme-UV (EUV; 100–900 Å) also photoionizes, heats and inflates the upper atmosphere, making it vulnerable to mass loss [31, 32]. Estimates of the total stellar insolation have been applied to planet evolution models to predict the end-state atmospheric properties of Neptune-sized and smaller planets [33, 34]. These planets can be left with a wide range of different radii and bulk densities, which are controlled both by the core masses and the mass of the remnant H/He-dominated envelope. Theoretical models predict a range of metal-enrichments, but it is not yet clear how the mass-loss process may alter these compositions. Currently there is very little observational data on the metallicity of these atmospheres. FINESSE observations of a large sample of planetary atmospheres around a wide variety of stars are uniquely capable of testing these models of thermal evolution and photoevaporative mass loss. Thus, FINESSE would explore the underlying causes of the mass-radius distribution by observing the behavior of atmospheric metallicity as a function of incident stellar flux for a given planet mass.

The diverse atmospheric properties of the planets observed by FINESSE would also provide us with clues as to how atmospheres evolved in our own Solar System. Some fraction of the FINESSE targets would be in compact, multiple-planet systems. These planets can gravitationally tug on each other as they orbit around the star, potentially leading to tidal heating of the planetary interiors. Alternatively, the mantles of close-in transiting planets could be heated by electromagnetic induction if the stellar magnetic field and rotation axes are inclined with respect to each other [35]. For gas-rich planets, this heating can affect the overall thermal structure and apparent radius during transit [e.g., 36]. For solid-surface planets, the heating could lead to volcanism, which in turn can profoundly influence the atmospheric composition and properties [e.g., 37]. Repeated episodes of volcanic outgassing and subsequent atmospheric escape could preferentially enrich the atmosphere in the heaviest elements, such as seemingly has occurred on Jupiter's moon Io, where sulfur dioxide dominates the composition after lighter hydrogen-bearing molecules and even carbon and nitrogen seemingly have escaped the moon over time [38]. Similarly, the Earth was thought to have been covered with a deep magma ocean early on in its history [39]. The properties of the thick steam and $CO_2$-rich atmosphere during that stage of terrestrial evolution affect how rapidly the Earth as a whole can shed the energy from planetary accretion and/or the moon-forming impact. FINESSE could potentially observe rocky Super Earths that may be hot enough to have molten silicate surfaces, which can result in exotic SiO, O, $H_2O$, $CO_2$, CO, and/or metal-rich atmospheres reminiscent of those in Earth's early history [e.g., 40, 41], thus illuminating processes that shaped the Earth's own atmosphere.

### 4. Conclusions

FINESSE, a NASA Medium-Explorer class mission currently in Step 2, provides the opportunity to place the study of exoplanet atmospheres on a statistical foundation. The large, uniform sample of exoplanet transmission and emission spectra FINESSE would obtain represents a tremendous scientific discovery opportunity for comparative planetology and in gaining deep insights into the incredibly diverse family of planets. FINESSE, building on the successes of Spitzer, Hubble, Kepler, and TESS, and being naturally synergistic with JWST, is a logical next step in studying exoplanets that we now know are common throughout the Galaxy.

### Acknowledgements

The information presented about FINESSE is pre-decisional and is provided for planning and discussion purposes only.

The content of this white paper is partially based on work done by the FINESSE science team, which includes Jacob Bean, Eliza Kempton, and Evgenya Shkolnik.

RTZ would like to thank Len Dorsky and Bill Frazier for their helpful comments.

This research has made use of the NASA Exoplanet Archive, which is operated by the California Institute of Technology, under contract with the National Aeronautics and Space Administration under the Exoplanet Exploration Program.

Part of the research was carried out at the Jet Propulsion Laboratory, California Institute of Technology, under contract with the National Aeronautics and Space Administration. © 2018. All rights reserved.